\documentclass[]{pasj01}
\draft

\begin{document} 
\Received{}
\Accepted{}

\title{Properties of the accretion ring in an X-ray binary, and accretion and excretion two-layer flows from it}

\author{Hajime \textsc{Inoue}\altaffilmark{1}}%
\altaffiltext{1}{Institute of Space and Astronautical Science, Japan Aerospace Exploration Agency, 3-1-1 Yoshinodai, Chuo-ku, Sagamihara, Kanagawa 252-5210, Japan}
\email{inoue-ha@msc.biglobe.ne.jp}


\maketitle

\begin{abstract}

We study properties of an accretion ring in a steady mass flow from a companion star to a compact object in an X-ray binary.
The accretion ring is a place where matter inflowing from a companion star sojourns for a while to bifurcate to accretion and excretion flows due to angular momentum transfer in it. 
The matter in the accretion ring rotates along the Keplerian circular orbit determined by the intrinsic specific angular momentum of the inflowing matter and forms a thick ring-envelope.
Two internal flows are expected to appear in the thick envelope.
One is a mass spreading flow bifurcating to a thick accretion flow and a thick excretion flow, as a result of the angular momentum transfer within the ring-envelope.
The other is a cooling flow toward the envelope center governed by radiative cooling under an effect of X-ray irradiation. 
This cooling flow eventually forms a core in the torus, from which a thin accretion disk and a thin excretion disk spread out as a result of the angular momentum transfer there again.
Evaluating and comparing the time scales for the two internal flows, the accretion ring is shown to generally originate a two-layer accretion flow in which a thin accretion disk is sandwiched by a thick accretion flow, unless the accretion rate is very low.
Properties of the thin excretion disk and the thick excretion flow are also investigated.
The thin excretion disk is expected to terminate at a distance 4 times as large as the accretion ring radius and to form another ring there, unless tidal effects from the companion star exist.  The thick excretion flow is, on the other hand, likely to turn to a super-sonic wind-flow reaching the infinity.

\end{abstract}

\section{Introduction}
It is widely accepted that matter inflowing from a companion star forms an accretion disk around a compact object in an X-ray binary.

Since the flow from the companion star moves in the rotational system of the binary and the Coriolis force induces its rotational motion, the inflowing matter can be approximated to have a constant specific angular momentum, $j_{0}$, and to tend to make a circularly rotating ring with a radius, $r_{0}$, defined as
\begin{equation}
r_{0} = \frac{j_{0}^{2}}{GM},
\label{eqn:r_0-Def}
\end{equation}
where $G$ and $M$ are the gravitational constant and the mass of the compact object respectively.
We call this radius the circularization radius (e.g. subsection 4.5 in Frank et al. 2002).

Pringle (1981) studied the time evolution of a ring having an initial mass distribution of the delta function at $r_{0}$, on an assumption that the specific angular momentum of the matter $j$ distributes as $j = \sqrt{rGM}$ for a position with distance $r$ from the gravitational center.
The result shows that the initial ring spreads out due to action of viscosity.
Almost all the mass moves inward losing energy and angular momentum, and the remaining small fraction of the matter moves out to larger radii in order to take up the angular momentum.

The time-dependent evolution of the accretion disks was generalized by taking account of interaction between the mass transfer stream from the companion star and the disk matter, and by introducing a parameter $\beta$ (by definition $0 \le \beta \le 1$) to describe the efficiency with which the stream is stripped and mixed with the disk matter, by Bath, Edwards \& Mantle (1983).
For $\beta \simeq 1$ (i.e. efficient stripping) the stream is stopped near the outer edge whereas for low $\beta$ it penetrates down to $r_{0}$ and forms a ring-like enhancement of the surface density.

In the scenario of the accretion disk evolution as shown above, the outer edge of the disk is considered to be generally at some radius exceeding the circularization radius $r_{0}$.  Almost all the angular momentum carried by the matter from the companion star is transported outward through the disk and is likely to be fed back into the binary orbit through tides exerted on the outer disk by the companion star (e.g. subsection 4.5 in Frank et al. 2002).

However, we have two arguments against the above picture.

\subsection{Survival time of accretion ring}
One is about the assumption of $j = \sqrt{rGM}$ independent of time adopted in the literature.

The basic equations for the present issue are 
the mass transfer equation
\begin{equation}
\frac{\partial \Sigma}{\partial t} = - \frac{1}{r} \frac{\partial}{\partial r}(r\Sigma u)
\label{eqn:MassCont}
\end{equation}
and
the angular momentum transfer equation
\begin{equation}
\frac{\partial}{\partial t} (\Sigma j) = -\frac{1}{r} \frac{\partial}{\partial r} (r\Sigma u j) + \frac{1}{r} \frac{\partial}{\partial r} ( r^{3}\Sigma \nu \frac{\partial \Omega}{\partial r}), 
\label{eqn:AM_Cont_Exact}
\end{equation}
where $\Sigma$, $u$, $\nu$ and $\Omega$ are the surface density, the radial velocity, the kinetic viscosity and the angular velocity of the disk respectively.
In the studies in the literatures, $(\partial j/\partial t)=0$ is assumed and equation (\ref{eqn:AM_Cont_Exact}) is replaced as
\begin{equation}
j\frac{\partial \Sigma}{\partial t} = -\frac{1}{r} \frac{\partial}{\partial r} (r\Sigma u j) + \frac{1}{r} \frac{\partial}{\partial r} (- \frac{3}{2} r^{2}\Sigma \nu \Omega_{\rm K}).
\label{eqn:AM_Cont}
\end{equation}
Then, by eliminating $u$ from equations (\ref{eqn:MassCont}) and (\ref{eqn:AM_Cont}), a diffusion equation for the surface density to evolve is obtained as
\begin{equation}
\frac{\partial \Sigma}{\partial t} = \frac{3}{r} \frac{\partial}{\partial r} \left \{ r^{1/2}
\frac{\partial}{\partial r} (\nu \Sigma r^{1/2}) \right \},
\label{eqn:DiffEq-Sigma}
\end{equation}
where $\Omega_{\rm K} = \sqrt{GM/r^{3}}$ is assumed, 
and this equation was solved by Pringle (1981).
If the size of the ring is $d$, the above equation indicates that the time scale for the ring to evolve, $t_{\rm e,\; Pringle}$, can be approximated as
\begin{equation}
t_{\rm e, \; Pringle} \sim \frac{d^{2}}{\nu} \sim \frac{d}{\alpha c_{\rm s}},
\label{eqn:t_s_Pringle}
\end{equation}
where we have introduced an approximate relation of $\nu$ with the viscous parameter $\alpha$, the sound velocity $c_{\rm s}$ and the relevant size $d$ as
\begin{equation}
\nu = \alpha c_{\rm s} d,
\label{eqn:nu-alpha}
\end{equation}
according to the $\alpha$-model for the viscosity (Shakura \& Sunyaev 1973).
Hence, the ring-evolution model by Pringle (1981) shows that the initial density enhancement of the ring quickly spreads out on the time scale longer than the dynamical time scale of $d/c_{\rm s}$ only by $\alpha^{-1}$.

However, the assumption of $(\partial j/\partial t)=0$ is not appropriate for the ring evolution because of the following reason.

From equations (\ref{eqn:MassCont}) and (\ref{eqn:AM_Cont_Exact}), we get
\begin{equation}
\Sigma \frac{\partial j}{\partial t} = -\Sigma u \frac{\partial j}{\partial r} + \frac{1}{r} \frac{\partial}{\partial r} ( r^{3}\Sigma \nu \frac{\partial \Omega}{\partial r}),
\label{eqn:AM_Change}
\end{equation}
which is neglected by setting $\partial j/\partial t = 0$ in the model by Pringle.
This equation is, however, very important, since such a ring as considered by Pringle  can start its spreading only after 
its angular momentum distribution changes from the initial condition, $j$ = constant  to the extension condition, $j \propto r^{1/2}$.
Since $j$ = constant initially, the first term on the right side of equation (\ref{eqn:AM_Change}) can be neglected and the time scale for the ring to start extension, $t_{\rm e}$, is approximately given as
\begin{equation}
t_{\rm e} \sim \frac{r d}{\nu} \sim \frac{r}{\alpha c_{\rm s}}.
\label{eqn:t_e}
\end{equation}
This time scale is longer than that in the Pringle's model by $r/d$ and independent of the initial ring size.
This means that the initial ring formed by the matter inflowing from the companion star should survive on a fairly long time before spreading out to two separated flows towards and outwards the compact object.  
We call this initial ring as the accretion ring and study its structure taking account of effects of X-ray irradiation in this paper.

\subsection{Outer boundary of steady accretion disk}\label{OuterBoundary}
The other argument is about the outer boundary of a steady accretion disk.

In the picture in the literatures, the stream from the companion star gradually merges with the outer part of the accretion disk and the outer edge of the disk exists at somewhere exceeding the circularization radius $r_{0}$.
Here, we define the effective outermost radius of the accretion disk $r_{\rm out}$ from which all the inflowing matter can be considered to flow inward through the disk.

From equations (\ref{eqn:MassCont}) and (\ref{eqn:AM_Cont_Exact}), we can have an angular momentum flow rate in a steady accretion disk as
\begin{equation}
-\dot{M} \sqrt{rGM} - 2\pi r^{3} \nu \Sigma \frac{\partial \Omega}{\partial r} = C_{\rm in},
\label{eqn:AM_inflow_rate}
\end{equation}
where $C_{\rm in}$ is the total angular momentum flow rate through a steady accretion disk and is given as
\begin{equation}
C_{\rm in} = -\dot{M} \sqrt{r_{\rm in}GM}.
\label{eqn:C_in}
\end{equation}
$\dot{M} = -2\pi r \Sigma u$ is the accretion rate through the accretion disk and $r_{\rm in}$ is the innermost radius of the disk.
Thus, we can say that 
\begin{equation}
-\left( 2\pi r^{3} \nu \Sigma \frac{\partial \Omega}{\partial r}\right)_{r = r_{\rm out}} \simeq \dot{M} \sqrt{r_{\rm out}GM},
\label{eqn:AMFR_r_out}
\end{equation}
must be transfered to some matter outside $r_{\rm out}$ via the viscous stress considering $C_{\rm in} \sim 0$ at $r_{\rm out}$ practically.
This indicates that the outflow rate of the angular momentum from the outermost edge of this accretion disk system is $\dot{M} \sqrt{r_{\rm out} GM}$, while the inflow rate from the companion star to the disk is $\dot{M} \sqrt{r_{0} GM}$.
Thus, as far as the steady state is considered, $r_{\rm out}$ should be $r_{0}$.
If so, it would be inevitable that a significant fraction of the inflowing matter from the companion star outflows from the accretion ring carrying the angular momentum transferred from the accreted matter through the disk.
The structure of the outflow from the ring is also studied in this paper.

\subsection{Overall picture of accretion ring}
\begin{figure}
 \begin{center}
  \includegraphics[width=12cm]{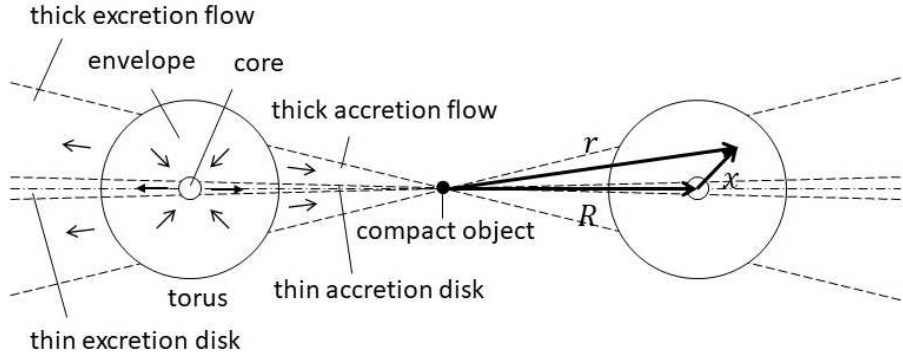} 
 \end{center}
\caption{Schematic diagram of a meridian cross section of an accretion ring.  For details, see text. }
\label{CrossSection}
\end{figure}

We briefly introduce an overall picture and its elements of the present study below.  
Names designated to respective elements in this paper are indicated in parentheses.
As a reference, a schematic view of the accretion ring and related elements are depicted in figure \ref{CrossSection}.

Matter inflowing from a companion star forms a fairly thin stream (the stream) as studied by Lubow \& Shu (1975; 1976).
The stream hits the accretion ring and merges with it. 
The matter from the stream just after the merging has such a high temperature as to make the ring cross section (the torus) thick.
A part of the matter in the outer part of the torus (the envelope) cools down and it induces an internal flow (the cooling flow) towards the center of the torus.
The matter through the cooling flow accumulates around the torus center and forms a low temperature core (the core).
Angular momentum is transferred from the inner half side facing to the gravity center (the compact object) to the outer half side on a time scale as given in equation (\ref{eqn:t_e}) and the ring starts spreading out.
As a result, a thin accretion disk (the accretion disk) and a thin excretion disk (the excretion disk) are expected to extend from the core, whereas a thick accretion flow (the thick accretion flow) and a thick excretion flow (the thick excretion flow) depart from the envelope.

Structures and properties of the ring and those of the excretion disk and flow are studied in sections 2, and 4 respectively, while some properties of the thick accretion flow is discussed in section 3. 
Comparisons with observations and some discussions are done in section 5.

\section{Properties of accretion ring}

\subsection{Ring formation}\label{RingFormation}
Initially, matter inflowing from a companion star is expected to form a ring with  
the circulation radius $r_{0}$.
Applying equation (4.21) in Frank, King \& Raine (2002) to several black hole binaries, $r_{0}$ is roughly estimated as
\begin{equation}
r_{0} \simeq 1 \times 10^{11} \mbox{ cm.}
\label{eqn:R}
\end{equation}
If the energy of the matter flowing in the stream from the companion star to the ring is conserved, we can approximately have the Bernoulli's equation for the stream as
\begin{equation}
\frac{v^{2}}{2} + \frac{5kT}{m_{\rm p}} - \frac{GM}{r} \simeq -\frac{1}{2} \frac{GM}{r_{\rm L1}},
\label{eqn:Bernoulli}
\end{equation}
on assumptions that the matter has the specific heat ratio of 5/3 and consists of fully ionized hydrogen.
Here, $v$, $T$,  $k$ and $m_{\rm p}$ are the velocity of the matter, its temperature, the Boltzmann constant and the proton mass respectively.  $r_{\rm L1}$ is a distance of the L1 point from the compact object and the factor 1/2 on the right sight is simply due to a rough assumption on the thermal energy at the L1 point. 
Since the matter is considered to rotate along the Keplerian circular orbit with the radius $r_{0}$ after landing on the accretion ring, the temperature of the ring matter, $T_{\rm in}$, just after the inflowing is calculated as
\begin{eqnarray}
T_{\rm in} &\simeq& \frac{1}{15} \frac{m_{\rm p} G M}{k r_{0}} \nonumber \\
&\simeq& 1.1 \times 10^{7} \left(\frac{M}{10 M_{\odot}}\right) \left(\frac{r_{0}}{10^{11} \mbox{ cm}}\right)^{-1} \mbox{ K},
\label{eqn:T_in}
\end{eqnarray}
assuming $r_{\rm L1} \simeq 3 r_{0}$.

\subsection{Hydro-static structure of the ring-tube}
We study structures of the torus on an assumption that all the matter in the torus has a common specific angular momentum, $j_{0}$, defined as
\begin{equation}
j_{0} = \sqrt{r_{0} GM},
\label{eqn:ell}
\end{equation}
and circularly rotates around the angular momentum axis keeping $j_{0}$ constant.
Here, we introduce a position vector, $\vec{x}$, from the torus center of a meridian cross section of the ring as
\begin{equation}
\vec{x} = \vec{r} - \vec{r_{0}},
\label{x_vec}
\end{equation}
where $\vec{r}$ is the position vector from the ring center and $\vec{r_{0}}$ is the position vector of the torus center from the ring center.
For the configuration, see a schematic diagram in figure \ref{CrossSection}.
We then recognize that if we neglect higher terms than the first order of $x/R$, the matter feels an axial-symmetrically attracting force in the anti $x$ direction from the center (Inoue 2012).
Thus, the mechanical structure on a meridian cross section of the torus can be approximated by a hydrostatic equation as
\begin{equation}
\frac{dP}{dx} = - n m_{\rm p} \frac{GMx}{r_{0}^{3}},
\label{eqn:dPdx}
\end{equation}
where $P$ and $n$ are the gas pressure and the number density of ions or atoms in the ring-matter.

Equation (\ref{eqn:dPdx}) can easily be solved on an isothermal approximation with the temperature, $T$, and 
the number density distribution is obtained as
\begin{equation}
n = n_{0} \exp \left(-\frac{x^{2}}{x_{\rm s}^{2}} \right),
\label{eqn:n}
\end{equation}
where $n_{0}$ is the number density at $x=0$ and $x_{\rm s}$ is the scale thickness of the torus given as
\begin{eqnarray}
x_{\rm s} &\simeq& \left( \frac{4 kT r_{0}^{3}}{m_{\rm p}GM} \right)^{1/2} \nonumber \\ 
&\simeq& 0.5 \left( \frac{T}{10^{7} \mbox{ K}}\right)^{1/2} \left(\frac{M}{10 M_{\odot}}\right)^{-1/2} \left(\frac{r_{0}}{10^{11} \mbox{ cm}}\right)^{1/2}r_{0}.
\label{eqn:x_s}
\end{eqnarray}

The total mass of the ring-tube, $M_{\rm r}$, is calculated as
\begin{eqnarray}
M_{\rm r} &\simeq& 2\pi r_{0} m_{\rm p} \int_{0}^{\infty} n 2\pi x dx \nonumber \\
&=& 2 \pi^{2} r_{0} x_{\rm s}^{2} n_{0} m_{\rm p}.
\label{eqn:M_r}
\end{eqnarray}
If we introduce the number density at $x = x_{\rm s}$ as the scale number density, $n_{\rm s} = e^{-1}n_{0}$, it is obtained from equations (\ref{eqn:M_r}) and (\ref{eqn:x_s}) as
\begin{eqnarray}
n_{\rm s} &\simeq& \frac{M_{\rm r}}{2\pi^{2} e r_{0} x_{\rm s}^{2} m_{\rm p}} \nonumber \\ &\simeq& \frac{M_{\rm r} GM}{8 \pi^{2} e r_{0}^{4} kT}.
\label{eqn:n_s}
\end{eqnarray}

As introduced in subsection 1.3, we consider that the torus consists of the envelope and the core in this study.
Then, the envelope and the core are approximated to both have the same density distribution as given in equation (\ref{eqn:n}) but to have different temperatures from each other.
Hereafter, the scale number density, temperature and total mass of the envelope and core are distinguished with the respective subscript ``e" and ``c".

\subsection{Ring-envelope}
\begin{figure}
 \begin{center}
  \includegraphics[width=12cm]{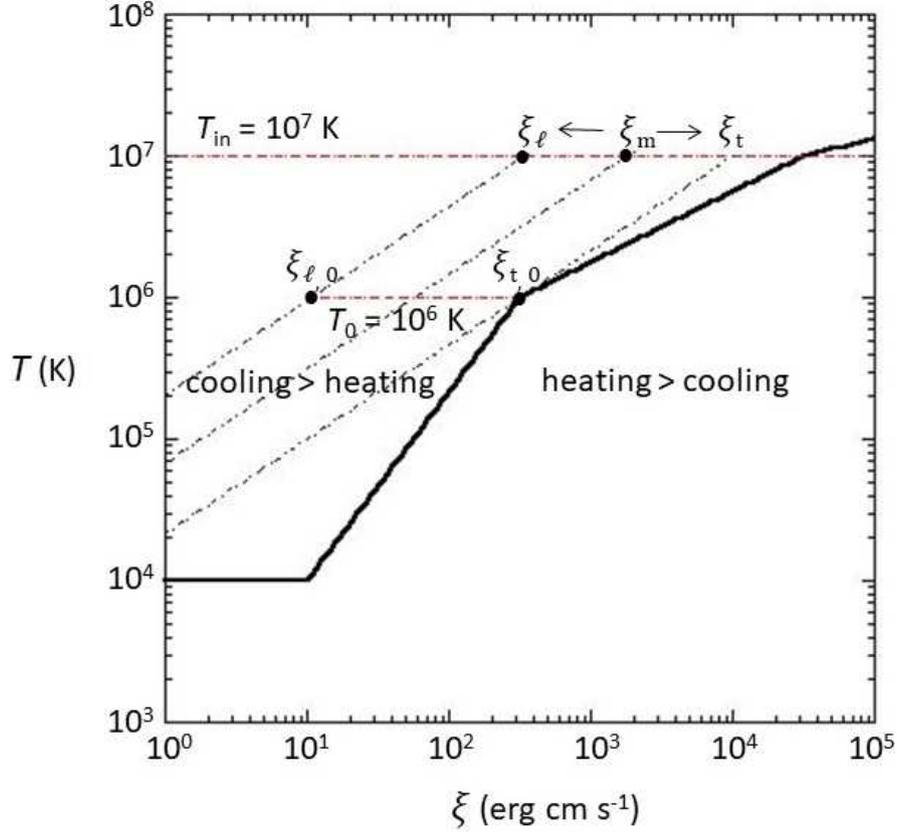} 
 \end{center}
\caption{Schematic theoretical $\xi$ - $T$ relation (thick zigzag line) and predicted $\xi$ - $T$ lines for three cases (dotted lines).  The upper dashed horizontal line shows the temperature, $T_{\rm in}$, and the lower dashed horizontal line indicates the envelope temperature, $T_{0}$, near $\xi_{\rm t,\; 0}$.  For details, see text. }\label{xi-T}
\end{figure}

The temperature of the envelope, $T_{\rm e}$, is basically $T_{\rm in}$ as given in equation (\ref{eqn:T_in}), but, more precisely, it deviates from it depending on an influence of X-ray irradiation in terms of the ionization parameter $\xi$ (Tarter et al. 1969) defined as
\begin{equation}
\xi = \frac{L}{nr^{2}},
\label{eqn:xi}
\end{equation}
where $L$ is an X-ray luminosity from the compact object, 
as discussed later.
In this paper, we adopt a schematic $\xi$ - $T$ relation as shown in figure \ref{xi-T}, which has been simplified from the result for the model 4 in Kallman \& McCray (1982).

The matter inflowing into the envelope stays there for certain time.
The staying time can be given by the shorter time scale between one for a mass spreading associated with angular momentum transfer and the other for a radiative  cooling under influence of X-ray irradiation.

\subsubsection{Mass spreading time scale}
Turbulent motions are likely to be always excited in the torus by the impact of the stream on it and to efficiently transfer the angular momentum from the inner half side facing to the compact object to the outer half side, causing the spreading of the torus matter.

Let typical scale and velocity of the turbulences be $\lambda$ and $\tilde{v}$, 
then the angular momentum transfer rate through the turbulent mixing, $\dot{J}_{\rm mix}$, per unit arc length of the boundary between the inner and outer half sides is approximately given as
\begin{equation}
\dot{J}_{\rm mix} \simeq - \left( \Sigma \tilde{v} \lambda r^{2} \frac{d\Omega}{dr}\right)_{r=r_{0}}.
\label{eqn:J_mix}
\end{equation}
As a result of the angular momentum transfer from the inner half side to the outer half side of a turbulent element, the specific angular momentum distribution tends to change from the initial one as $j_{0} = \sqrt{r_{0}GM}$ to that for the mass spreading as $j_{\rm s} = \sqrt{rGM}$.
Then, when $j$ becomes slightly flatter than $j_{\rm s}$, the mass spreading  begins; the matter in the inner half side starts to flow inward to the compact object, while that in the outer half side goes outward.
If we express the angular momentum flow rates from the inner side and the outer side as $\dot{J}_{\rm in}$ and $\dot{J}_{\rm out}$ respectively, they can be written as
\begin{equation}
\dot{J}_{\rm out} \simeq \Sigma u j_{0},
\label{eqn:J_out}
\end{equation}
and
\begin{equation}
\dot{J}_{\rm in} \simeq - \Sigma u j_{0},
\label{eqn:J_in}
\end{equation}
where $u$ is the absolute value of the flowing velocity which is assumed to be the same for the both sides.
Since the difference of the angular momentum flow rate, $\Delta \dot{J}$, between the outer side and the inner side defined as
\begin{equation}
\Delta \dot{J} = \dot{J}_{\rm out} - \dot{J}_{\rm in} = 2\Sigma u j_{0}
\label{eqn:Delta_J-dot}
\end{equation}
should be the result of the angular momentum transfer through the turbulent mixing, 
we have an equation as
\begin{equation}
\Delta \dot{J} = \dot{J}_{\rm mix}.
\label{eqn:Delta_J-J_mix}
\end{equation}
Inserting equations (\ref{eqn:J_mix}) and (\ref{eqn:Delta_J-dot}) into equation (\ref{eqn:Delta_J-J_mix}), and with the help of a relation as $\Omega = j_{0}/r^{2}$, we get 
\begin{equation}
u \simeq \frac{\tilde{v}\lambda}{r_{0}} \simeq \frac{\alpha c_{\rm s,\; e} x_{\rm s,\; e}}{r_{0}},
\label{eqn:u}
\end{equation}
where we have introduced the concept of the $\alpha$ model for the turbulent viscosity as $\tilde{v} \lambda \simeq \alpha c_{\rm s,\; e} x_{\rm s,\; e}$ in terms of the sound velocity $c_{\rm s,\; e}$ and the scale thickness $x_{\rm s,\; e}$ of the envelope.

The spreading flow rate $\dot{M}_{\rm s, \; e}$ to both the inner and outer sides can be approximated as
\begin{equation}
\dot{M}_{\rm s, \; e} \simeq 2 \times 2\pi r_{0} x_{\rm s,\; e} n_{\rm s,\; e} m_{\rm p} u.
\label{eqn:Mdot_s,e}
\end{equation}
Then, the time scale for the mass spreading, $t_{\rm s,\; e}$, can be estimated from equations (\ref{eqn:M_r}) and (\ref{eqn:Mdot_s,e}) as
\begin{eqnarray}
t_{\rm s, \; e} &\simeq& \frac{M_{\rm e}}{\dot{M}_{\rm s,\; e}} \nonumber \\ &\simeq& \frac{\pi e r_{0}}{2 \alpha}\left( \frac{2 kT_{\rm e}}{m_{\rm p}}\right)^{-1/2} \nonumber \\
&\simeq& 1.1 \times 10^{5}  \left(\frac{\alpha}{0.1}\right)^{-1} \left(\frac{r_{0}}{10^{11}\; \mbox{cm}}\right) \left(\frac{T_{\rm e}}{10^{7}\; \mbox{K}}\right)^{-1/2} \mbox{ s},
\label{eqn:t_s,e-estimated}
\end{eqnarray}
where $c_{\rm s,\; e}$ is expressed as
\begin{equation}
c_{\rm s,\; e} = \left( \frac{2 kT_{\rm e}}{m_{\rm p}} \right)^{1/2}.
\label{eqn:c_s}
\end{equation}

\subsubsection{Cooling time scale}
The typical cooling time scale of the envelope, $t_{\rm c,\; e}$, can be estimated as
\begin{equation}
t_{\rm c,\; e} \simeq \frac{\varepsilon 3 kT_{\rm e}}{n_{\rm s,\; e}\Lambda}
\label{eqn:t_c,e}
\end{equation}
where $\varepsilon$ ($\ge 1$) is a parameter expressing an effect of X-ray irradiation, 
which is given in appendix, and $\Lambda$ is the cooling function (e.g. Sutherland \& Dopita 1993).
When $t_{\rm c,\; e} \ll t_{\rm s,\; e}$, the total mass of the envelope, $M_{\rm e}$ is written as
\begin{equation}
M_{\rm e} \simeq \dot{M}_{0} \ t_{\rm c,\; e},
\label{eqn:M_e-Mdot}
\end{equation}
where $\dot{M}_{0}$ is the mass inflow rate from the companion star.
Since $M_{\rm e}$ relates to $n_{\rm s,\; e}$ as given in equation (\ref{eqn:M_r}), 
$t_{\rm c,\; e}$ can be calculated from equations (\ref{eqn:t_c,e}), (\ref{eqn:M_e-Mdot}) and (\ref{eqn:M_r}) as
\begin{eqnarray}
t_{\rm c, \; e} &\simeq& \left( \frac{24 \pi^{2} e \varepsilon}{\dot{M}_{0} GM \Lambda} \right)^{1/2} kT_{\rm e} r_{0}^{2} \nonumber \\
&\simeq& 1.9 \times 10^{4} \varepsilon^{1/2} \left( \frac{\Lambda}{\Lambda_{0} } \right)^{-1/2} \left( \frac{M}{10} \right)^{-1/2} \left( \frac{r_{0}}{10^{11}} \right)^{2} \left( \frac{T_{\rm e}}{10^{7}} \right) \left( \frac{\dot{M}_{0}}{10^{16}} \right)^{-1/2} \mbox{ s}.
\label{eqn:t_c_value}
\end{eqnarray}
where denominator numbers in parentheses for $M$, $R$, $T_{\rm e}$ and $\dot{M}_{0}$ are in units of erg cm$^{3}$ s$^{-1}$, $M_{\odot}$, cm, K and g s$^{-1}$ respectively and 
$\Lambda_{0} = 10^{-22.6}$ erg cm$^{3}$ s$^{-1}$ which is obtained at $T = 10^{7}$ K in the case of the solar abundance from Sutherland \& Dopita (1993).
We use the same units for the parameters in parentheses and omit the respective units in the following equations.

As seen from equation (\ref{eqn:t_c_value}), the cooling time $t_{\rm c,\; e}$ decreases as the mass inflow rate $\dot{M}_{0}$ increases.
Then, when $t_{\rm c,\; e}$ is shorter than the matter circulation time in the ring, $t_{\rm circ}$ defined as
\begin{eqnarray}
t_{\rm circ} &=& 2\pi \sqrt{\frac{r_{0}^{3}}{GM}} \nonumber \\ 
&=& 5.4 \times 10^{3} \left( \frac{M}{10} \right)^{-1/2} \left( \frac{r_{0}}{10^{11}} \right)^{3/2} \mbox{ s},
\label{eqn:t_circ}
\end{eqnarray}
the thick envelope does not appear from the beginning.
However, the effect of the X-ray heating does not allow such a situation as $t_{\rm c,\; e} < t_{\rm circ}$ to happen and ensures that the thick envelope should always exist, as discussed below.

Figure \ref{xi-T} shows that when $\xi \gtrsim 10^{4}$ erg cm s$^{-1}$, the X-ray heating makes a temperature of an illuminated matter as high as $\sim 10^{7}$ K.
We can relate the X-ray luminosity from a compact object $L$ to the mass inflow rate $\dot{M}_{0}$ as
\begin{equation}
L \simeq \eta \frac{\dot{M}_{0}}{2} c^{2},
\label{eqn:L-M_0}
\end{equation}
where $\eta$ is the energy conversion efficiency associated with the mass accretion onto the compact object and the factor 1/2 multiplied to $\dot{M}_{0}$ comes from an assumption that a half of the inflowing matter is accreted by the compact star.
$n_{\rm s, \; e}$ is, on the other hand, given as in equation (\ref{eqn:n_s}) and $M_{\rm e}$ relates with $\dot{M}_{0}$ as $M_{\rm e} = \dot{M}_{0} \ t_{\rm stay, \; e}$ in terms of the matter staying time $t_{\rm stay,\; e}$ in the envelope.
Hence, we can get the following condition on $t_{\rm stay, \; e}$ for the sufficient X-ray heating to occur from an equation $L/(n_{\rm s,\; e} r_{0}^{2}) \gtrsim \xi_{4} = 10^{4}$ erg cm s$^{-1}$ as
\begin{eqnarray}
t_{\rm stay,\; e} &\lesssim& \frac{4\pi^{2} \eta c^{2} e r_{0}^{2} k T_{\rm e}}{\xi_{4} GM} \nonumber \\ &\simeq& 1.0 \times 10^{4} \left( \frac{\eta}{0.1 } \right) \left( \frac{M}{10} \right)^{-1} \left( \frac{r_{0}}{10^{11}} \right)^{2} \left( \frac{T_{\rm e}}{10^{7}} \right)  \mbox{ s}. 
\label{eqn:t_stay-Cond}
\end{eqnarray}
This means that X-ray heating prevents a situation as $t_{\rm c,\; e} < t_{\rm circ}$ from appearing by enlarging $\varepsilon$ to much larger than unity for parameter values considered here.

\subsubsection{Internal flows in the envelope}
Following the above discussions, the envelope is expected to have two internal flows.
One is a flow towards the torus center as a result of the radiative cooling, and the other is a flow spreading out from the torus, bifurcating to a thick accretion flow and a thick excretion flow, as a result of the angular momentum transfer.
We call the former and the latter flows respectively as the cooling flow and the mass-spreading flow hereafter.

The mass flow rates through the cooling flow and the mass-spreading flow, $\dot{M}_{\rm c, \; e}$ and $\dot{M}_{\rm s,\; e}$, are approximately calculated as
\begin{equation}
\dot{M}_{\rm c,\; e} = \frac{M_{\rm e}}{t_{\rm c,\; e}},
\label{eqn:Mdot_c,e-Me}
\end{equation}
and
\begin{equation}
\dot{M}_{\rm s,\; e} = \frac{M_{\rm e}}{t_{\rm s,\; e}},
\label{eqn:Mdot_s,e-Me}
\end{equation}
where $t_{\rm c,\; e}$ and $t_{\rm s,\; e}$ are given in equations (\ref{eqn:t_s,e-estimated}) and (\ref{eqn:t_c_value}) respectively.
In the steady state, the following equation should be satisfied as
\begin{equation}
\dot{M_{0}} = \dot{M}_{\rm c,\; e} + \dot{M}_{\rm s,\; e}.
\label{eqn:Mdot_0-Sum}
\end{equation}

From the above three equations (\ref{eqn:Mdot_c,e-Me}), (\ref{eqn:Mdot_s,e-Me}) and (\ref{eqn:Mdot_0-Sum}), we get 
\begin{equation}
M_{\rm e} = \frac{t_{\rm c,\; e} t_{\rm s,\; e}}{t_{\rm c, \; e} + t_{\rm s,\; e}} \dot{M}_{0},
\label{eqn:M_e-Mdot0}
\end{equation}
\begin{equation}
\dot{M}_{\rm c,\; e} = \frac{t_{\rm s, \; e}}{t_{\rm c,\; e}+t_{\rm s,\; e}} \dot{M}_{0},
\label{eqn:Mdot_c,e-Mdot0}
\end{equation}
and
\begin{equation}
\dot{M}_{\rm s,\; e} = \frac{t_{\rm c, \; e}}{t_{\rm c,\; e}+t_{\rm s,\; e}} \dot{M}_{0}.
\label{eqn:Mdot_s,e-Mdot0}
\end{equation}

\subsection{Ring core}
Since the destination of the cooling flow is the torus center, the matter through the cooling flow should accumulate there.
The cooling function sharply decreases as the temperature decreases to $\sim 10^{4}$ K or lower, and thus the matter is expected to form a core with the temperature $T_{\rm c} \sim 10^{4}$ K around the torus center.
The scale thickness of the core, $x_{\rm s, \; c}$ is estimated from equation (\ref{eqn:x_s}) by setting $T$ as $T_{\rm c}$ to be $\sim 0.02\ r_{0}$ for $T_{\rm c} = 10^{4}$ K and $r_{0} = 10^{11}$ cm.

Angular momentum transfer should happen even in the core and the mass is expected to spread out to a thin accretion disk and a thin excretion disk.
The mass spreading time of the core, $t_{\rm s,\; c}$, can be calculated, similarly to the case of equation (\ref{eqn:t_s,e-estimated}), as
\begin{eqnarray}
t_{\rm s, \; c}  &\simeq& \frac{\pi e r_{0}}{2 \alpha}\left( \frac{ kT_{\rm c}}{m_{\rm p}}\right)^{-1/2} \nonumber \\
&\simeq& 6.7 \times 10^{6}  \left(\frac{\alpha}{0.1}\right)^{-1} \left(\frac{r_{0}}{10^{11}\; \mbox{cm}}\right) \left(\frac{T_{\rm c}}{10^{4}\; \mbox{K}}\right)^{-1/2} \mbox{ s},
\label{eqn:t_s,c-estimated}
\end{eqnarray}
where the hydrogen gas is assumed to be neutral.

\subsection{Effects of X-ray irradiation on the envelope}\label{X-rayIrradiation}

\subsubsection{Case of low accretion rate}\label{LowAccretionRateCase}

By comparing equations (\ref{eqn:t_c_value}) and (\ref{eqn:t_s,e-estimated}), 
we see that $t_{c, \; e} \gg t_{s, \; e}$ when $\dot{M}_{0} \ll \dot{M}_{1}$. $\dot{M}_{1}$ is given by equating $t_{\rm c,\; e}$ with $t_{\rm s,\; e}$ as
\begin{equation}
\dot{M}_{1} \simeq 3.3 \times 10^{14} \left(\frac{\alpha}{0.1} \right)^{2} 
 \left( \frac{\Lambda}{\Lambda_{0} } \right)^{-1} \left( \frac{M}{10} \right)^{-1} \left( \frac{r_{0}}{10^{11}} \right)^{2} \left( \frac{T_{\rm e}}{10^{7}} \right)^{3} \mbox{ g s$^{-1}$},
\label{eqn:M_1}
\end{equation}
where $\varepsilon$ is set to be 1, since X-ray heating is negligible here.
In this case, it is expected that all the inflow from the companion star is transferred to the thick spreading flow, namely  $\dot{M}_{\rm s,\; e} \simeq \dot{M}_{0}$ and $\dot{M}_{\rm c, \; e} \simeq 0$.

If $\dot{M}_{\rm s,\; e}$ is evenly divided into the accretion flow and the excretion flow, the accretion rate from the envelope, $\dot{M}_{\rm ac,\; e}$, is expressed as
\begin{equation}
\dot{M}_{\rm ac,\; e} \simeq \frac{\dot{M}_{\rm s,\; e}}{2} \simeq \frac{\dot{M}_{0}}{2},
\label{eqn:Mdot_ac,e-Mdot_s,e}
\end{equation}
and
\begin{equation}
L \simeq \eta \frac{\dot{M}_{0}}{2} c^{2}.
\label{eqn:L-case_L}
\end{equation}
Since $M_{\rm e} \simeq t_{\rm s, \; e} \dot{M}_{0}$ in this case, the scale number density, $n_{\rm s,\; e}$, is estimated from equation (\ref{eqn:n_s}) as
\begin{equation}
n_{\rm s,\; e} \simeq \frac{t_{\rm s, \; e} \dot{M}_{0} GM}{8\pi^{2} e r_{0}^{4} kT_{\rm e}}.
\label{eqn:n_s-case_L}
\end{equation}
Then, the ionization parameter, $\xi_{\ell}$, in this case, is calculated with the help of equation (\ref{eqn:t_s,e-estimated}) as
\begin{eqnarray}
\xi_{\ell} &\simeq& \frac{4\pi^{2} \eta c^{2} e r_{0}^{2} kT_{\rm e}}{t_{\rm s,\; e} GM } \nonumber \\
&\simeq& \frac{4\pi \eta c^{2} \alpha r_{0} (kT_{\rm e})^{3/2}}{GM m_{\rm p}^{1/2}} \nonumber \\
&\simeq& 3.4 \times 10^{2} \left( \frac{\eta}{0.1} \right) \left( \frac{\alpha}{0.1} \right) \left( \frac{M}{10} \right)^{-1} \left( \frac{r_{0}}{10^{11}} \right) \left( \frac{T_{\rm e}}{10^{7}} \right)^{3/2} \mbox{ erg cm s$^{-1}$}.
\label{eqn:xi_ell}
\end{eqnarray}

This $\xi_{\ell}$ - $T_{\rm e}$ line is drawn in figure \ref{xi-T} and is seen to be in the range where the effect of X-ray heating is negligibly small.
Even if the X-ray heating is negligible, the density is too low for the envelope to radiatively cool down and the ring tube is expected to be kept thick with the initial temperature, $T_{\rm in}$, in the low-accretion-rate case.   And, a thick accretion flow and a thick excretion flow are expected to extend from the envelope.

\subsubsection{Case of medium accretion rate}
When $\dot{M}_{0} \gtrsim \dot{M}_{1}$, $t_{\rm c, \; e}$ comes to satisfy the relation as
$t_{\rm c.\; e} \lesssim t_{\rm s,\; e}$ and the mass inflow rate, $\dot{M}_{0}$, from the companion star is divided into the two components, $\dot{M}_{\rm c,\; e}$ and $\dot{M}_{\rm s,\; e}$ following equations (\ref{eqn:Mdot_c,e-Mdot0}) and (\ref{eqn:Mdot_s,e-Mdot0}) respectively.
If we assume again that $M_{\rm c,\; e}$ turns evenly to the thin accretion disk and the thin excretion disk from the core in the steady state, the accretion rate from the core, $\dot{M}_{\rm ac,\; c}$, is written as
\begin{equation}
\dot{M}_{\rm ac,\; c} \simeq \frac{\dot{M}_{\rm c,\; e}}{2}.
\label{eqn:Mdot_c,a-Mdot_e,c}
\end{equation}
The total X-ray luminosity is reasonably expressed with the help of equations (\ref{eqn:Mdot_c,e-Mdot0}) and (\ref{eqn:Mdot_s,e-Mdot0}) as
\begin{eqnarray}
L &\simeq& \eta \; \left( \frac{\dot{M}_{\rm c, \; e}}{2} + \frac{\dot{M}_{\rm s, \; e}}{2} \right) \; c^{2} \nonumber \\ &=& \eta \; \frac{\dot{M}_{0}}{2} \; c^{2}.
\label{eqn:L-caseM}
\end{eqnarray}
The scale number density of the envelope in this case is calculated from equations (\ref{eqn:n_s}) and (\ref{eqn:M_e-Mdot0}) as
\begin{equation}
n_{\rm s,\; e} \simeq \frac{t_{\rm c,\; e}t_{\rm s,\; e}}{t_{\rm c,\; e}+t_{\rm s,\; e}} \dot{M}_{0} \frac{GM_{\rm X}}{8\pi^{2} e r_{0}^{4} kT_{\rm e}}.
\label{eqn:n_s-caseM}
\end{equation}
Then, the ionization parameter, $\xi_{\rm m}$, is given with the help of equation (\ref{eqn:xi_ell}) as
\begin{eqnarray}
\xi_{\rm m} &\simeq& \frac{t_{\rm c,\; e}+t_{\rm s,\; e}}{t_{\rm c,\; e}t_{\rm s,\; e}} \frac{4\pi^{2} e \eta c^{2} r_{0}^{2} kT_{\rm e}}{GM} \nonumber \\
&\simeq& \frac{t_{\rm c,\; e}+t_{\rm s,\; e}}{t_{\rm c,\; e}} \; \xi_{\ell}.
\label{eqn:xi_m}
\end{eqnarray}

In this medium accretion rate range, as $\dot{M}_{0}$ increases from $\dot{M}_{1}$, $t_{\rm c, \; e}$  decreases according to its proportionality to $\dot{M}_{0}^{-1/2}$ as in equation (\ref{eqn:t_c_value}), and $\xi_{\rm m}$ increases due to the $(t_{\rm c,\; e}+t_{\rm s,\; e})/t_{\rm c,\; e}$ term in equation (\ref{eqn:xi_m}).
Then, the $\xi_{\rm m}$ line moves right in figure \ref{xi-T} and eventually gets close to the line tagged with $\xi_{\rm t}$ which touches the theoretical $\xi$ - T curve at $\xi_{\rm t,\; 0}$.
Here, we define $\xi_{\rm t}$ as
\begin{equation}
\xi_{\rm t} \simeq 1 \times 10^{4} \left( \frac{T_{\rm e}}{10^{7}} \right)^{3/2} \mbox{ erg cm s$^{-1}$}.
\label{eqn:xi_t}
\end{equation}
If $\xi_{\rm m}$ just reaches $\xi_{\rm t,\; 0}$, the envelope should be governed by the situation at the tangential point where the radiative cooling rate balances with the X-ray heating rate, and then, $t_{\rm c,\; e}$ should be regarded to be infinity as discussed in appendix.
Since this situation of $t_{\rm c,\; e} \gg t_{\rm s,\; e}$ is the same as discussed in the low accretion rate case in sub-subsection \ref{LowAccretionRateCase}, $\xi$ is required to return to $\xi_{\ell}$ then.
This means that $\xi$ of the envelope cannot get just to $\xi_{\rm t}$, and a steady state solution exists in the vicinity of $\xi_{\rm t}$, which will be discussed in the next sub-subsection.

If we define the medium accretion rate range as in the case of $\dot{M}_{0} \gtrsim \dot{M}_{1}$ and $\xi_{\rm m} \ll \xi_{\rm t}$, the upper bound of $\dot{M}_{0}$ in this range, $\dot{M}_{2}$, is calculated by equating $\xi_{\rm m}$ in equation (\ref{eqn:xi_m}) to $\xi_{\rm t}$ in equation (\ref{eqn:xi_t}) with the help of equation (\ref{eqn:t_c_value}) and setting $\varepsilon = 1$ as
\begin{equation}
\dot{M}_{2} \simeq 3.7 \times 10^{16} \left( \frac{\eta}{0.1} \right)^{-2} \left( \frac{\Lambda}{\Lambda_{0}}\right)^{-1} \left( \frac{M}{10} \right) \left( \frac{T_{\rm e}}{10^{7}} \right)^{3} \mbox{ g s$^{-1}$}.
\label{eqn:Mdot_2}
\end{equation}

In the medium accretion rate range with $\dot{M}_{1} \lesssim \dot{M}_{0} \ll \dot{M}_{2}$, the cooling time becomes comparable to or shorter than the mass spreading time.  Although the effect of the X-ray heating is not so significant yet, the temperature of the envelope is still kept $\sim T_{\rm in}$ during the cooling time.  Then, two parallel accretion flows are expected to appear.
One is a thin disk accretion flow from the core with the accretion rate, $\dot{M}_{\rm ac,\; c}$, as
\begin{equation}
\dot{M}_{\rm ac,\; c} \simeq F \frac{\dot{M}_{0}}{2},
\label{eqn:Mdot_ac,a-caseM}
\end{equation}
and the other is a thick accretion flow from the envelope with the accretion rate, $\dot{M}_{\rm ac,\; e}$, as
\begin{equation}
\dot{M}_{\rm ac,\; e} \simeq (1-F) \frac{\dot{M}_{0}}{2}.
\label{eqn:Mdot_ac,e-caseM}
\end{equation}
Here, $F$ is defined as
\begin{equation}
F = \frac{t_{\rm s,\; e}}{t_{\rm c,\; e}+t_{\rm s,\; e}},
\label{eqn:F}
\end{equation}
where $t_{\rm c,\; e}$ is a function of $\dot{M}_{0}$ as in equation (\ref{eqn:t_c_value}).
In this range, $t_{\rm c,\; e} \lesssim t_{\rm s, \; e}$ and thus $F$ is in a range of $1/2 \lesssim F < 1$.
The accretion rate through the thin accretion disk is slightly dominant to that through the thick accretion flow in this case.

\subsubsection{Case of high accretion rate}\label{HighAccretionRateCase}
When $\dot{M}_{0} \gtrsim \dot{M}_{2}$, $\xi$ of the envelope gets close to $\xi_{\rm t,\; 0}$ and the effect of the X-ray irradiation on the envelope becomes significant.
The ionization parameter in this $\dot{M}_{0}$ range, $\xi_{\rm h}$, is determined, utilizing equations (\ref{eqn:t_c}) and (\ref{varepsilon}) in appendix, by an equation as
\begin{eqnarray}
\xi_{\rm h} &\simeq& \frac{t_{\rm c,\; e}+t_{\rm s,\; e}}{t_{\rm c,\; e}} \; \xi_{\ell, \; 0} \nonumber \\ &=& \left[ 1 + \frac{t_{\rm s,\; e}}{t_{\rm c, \; e,\; 0}} \left( 1-\frac{\xi_{\rm h}}{\xi_{\rm t,\; 0}} \right)^{1/2} \right] \xi_{\ell. \; 0}
\label{eqn:xi_h}
\end{eqnarray}
to which equation (\ref{eqn:xi_m}) is modified as a $\xi$ - $T$ relation along the $T_{0} = 10^{6}$ K line between $\xi_{\ell,\; 0}$ and $\xi_{t,\; 0}$ as revealed in figure \ref{xi-T}.
When $\dot{M}_{0}$ exceeds $\dot{M}_{2}$, $\xi_{\rm h}$ should have become larger than $\xi_{\rm t,\; 0}$ unless X-ray heating exist, but the term of $[1 - (\xi_{\rm h}/\xi_{\rm t,\; 0})]^{1/2}$ reflecting the effect of X-ray heating suppresses $\xi_{\rm h}$ slightly lower than $\xi_{\rm t,\; 0}$.

Combining equations (\ref{eqn:F}) and (\ref{eqn:xi_h}) and considering $\xi_{\rm h} \sim \xi_{\rm t,\; 0}$, we have
\begin{eqnarray}
F &\simeq& 1\ -\ \frac{\xi_{\ell,\; 0}}{\xi_{\rm h}} \nonumber \\ &\sim& 1\ -\ X
\label{eqn:F-xi_h}
\end{eqnarray}
where $X$ is defined and calculated with equations (\ref{eqn:xi_ell}) and (\ref{eqn:xi_t}) as
\begin{eqnarray}
X &=& \frac{\xi_{\ell,\; 0}}{\xi_{\rm t,\; 0}} \nonumber \\ 
&\simeq& 3 \times 10^{-2} \left( \frac{\eta}{0.1} \right) \left( \frac{\alpha}{0.1} \right) \left( \frac{M}{10} \right)^{-1} \left( \frac{r_{0}}{10^{11}} \right).
\label{eqn:X}
\end{eqnarray}

In this high accretion rate range, the accretion rate through the thick flow is considered to be approximately several \% as small as that through the thin disk.

\subsection{Impact of the stream on the accretion ring}
The properties of the accretion ring in the steady state have been studied above.
Here, we confirm that the stream from the companion star does not give so large impact on the accretion ring as to conflict with the above studies when it merges with the ring.

\subsubsection{Impact on the core}

The ram pressure of the stream on the surface of the torus, $P_{\rm in}$, can be obtained as
\begin{equation}
P_{\rm in} \simeq \rho v_{\bot}^{2},
\label{eqn:ram_pressure}
\end{equation}
where $\rho$ is the density of the stream and $v_{\bot}$ is its velocity component in the direction perpendicular to the torus surface when it hits the torus.
Since we can approximate the mass flow rate of the stream, $\dot{M}_{0}$ as
\begin{equation}
\dot{M}_{0} \simeq \rho v_{\rm in} \pi d^{2},
\label{eqn:M_0}
\end{equation}
where $v_{\rm in}$ is the stream velocity and $d$ is the radius of the stream-cross section, $P_{\rm in}$ can be rewritten as
\begin{equation}
P_{\rm in} \simeq \frac{\dot{M}_{0}}{\pi d^{2}} \frac{v_{\bot}^{2}}{v_{\rm in}}.
\label{eqn:P_in}
\end{equation}

If the orbital radius of the matter in the ring shifts $\Delta r$ from the Keplerian circular motion by the ram pressure of the stream, the repulsive force per mass, $f$, can be calculated as
\begin{equation}
f \simeq \frac{GM}{r_{0}^{2}} \frac{\Delta r}{r_{0}}.
\label{eqn:f}
\end{equation}
Since the average column density of the ring-core in the radial direction is roughly given as $M_{\rm c}/(4\pi r_{0}  x_{\rm s,\; c})$, the repulsive pressure of the ring-core, $P_{\rm rep}$, can be approximated as
\begin{equation}
P_{\rm rep} \simeq \frac{M_{\rm c}}{4\pi r_{0} x_{\rm s,\; c}} \frac{GM}{r_{0}^{2}}\frac{\Delta r}{r_{0}}.
\label{eqn:P_rep}
\end{equation}

By equating $P_{\rm in}$ to $P_{\rm rep}$, we have a relation
\begin{equation}
\frac{\Delta r}{x_{\rm s,\; c}} \simeq \frac{2}{\pi} \frac{v_{\bot}^{2}}{v_{\rm K}v_{\rm in}} \frac{t_{\rm circ}}{t_{\rm s,\; c}} \left(\frac{d}{r_{0}}\right)^{-2},
\label{eqn:Dr/x_s}
\end{equation}
where $M_{\rm c} = \dot{M}_{0} t_{\rm s,\; c}$ is assumed and $v_{\rm K} = \sqrt{GM/r_{0}}$.
Considering $t_{\rm circ}/t_{\rm s,\; c} \sim 10^{-3}$ as seen from equations (\ref{eqn:t_circ}) and (\ref{eqn:t_s,c-estimated}) for appropriate parameters, and $v_{\bot} < v_{\rm K} < v_{\rm in}$, a moderate shift with $\Delta r < x_{\rm s,\; c}$ can make the ring-core stop the stream unless $d$ is much less than a few \% of $r_{0}$.
Since $d$ could be several \% or a little more of $r_{0}$ (e.g. Lubow \& Shu 1975; 1976), the stream can be considered to be well braked by the core. 

\subsubsection{Effect on the hot envelope}
Once the stream stops around the core surface, the matter in the stream should be heated up to $T_{\rm in} \sim 10^{7}$ K as discussed in subsection \ref{RingFormation}.
Then, it is expected to expand to the same size as the scale thickness of the envelope, $x_{\rm s,\; e}$, since the temperature of the envelope is considered to be the same as $T_{\rm in}$.
A key question is whether the expanded matter cools down soon or not.

The expanded matter of the stream just after the hit to the core should start flowing in the envelope, with velocity $v_{\rm K}$ and with mass flow rate $\dot{M}_{0}$ in the steady state.
If we assume that the flow of the fresh matter from the stream has eventually the same cross section as the ring-envelope, the average number density of the fresh flow, $n_{\rm f}$, can be written from the continuity equation as
\begin{equation}
n_{\rm f} \simeq \frac{\dot{M}_{0}}{\pi m_{\rm p} x_{\rm s,\; e}^{2} v_{\rm K}}.
\label{eqn:n_f}
\end{equation}
On the other hand, the scale number density of the envelope, $n_{\rm s,\; e}$, is expressed in equation (\ref{eqn:n_s}) replacing $M_{\rm r}$ to $M_{\rm e}$ in equation (\ref{eqn:M_e-Mdot}).

From the above equations we get
\begin{equation}
\frac{n_{\rm f}}{n_{\rm s,\; e}} \simeq e \frac{t_{\rm circ}}{t_{\rm c,\; e}}.
\label{eqn:n_f-n_s,e}
\end{equation}
This equation indicates that $n_{\rm f}$ should be smaller than $n_{\rm s,\; e}$ under a situation that the cooling flow time scale of the ring envelope is sufficiently longer than the rotational period of the ring.
This situation is the premise of the present sturdy.  Otherwise, the geometrically thick ring should not have been realized from the beginning.
Thus, considering that the time scale of the radiative cooling is inversely proportional to the number density, we cay say that the cooling time of the hot matter newly added from the stream to the ring-envelope should be longer than that in the main flow, as far as the premise is held.  

The impact of the stream on the ring can be considered to give no large effect on the basic frame of the present study.

\section{Two-layer accretion flow}
As discussed in subsection \ref{X-rayIrradiation}, the accretion ring is expected to generate the two-layer accretion flow consisting of the thin accretion disk and the thick accretion flow, but the dividing ratio of the accretion rate into the two components changes with $\dot{M}_{0}$.

When $\dot{M}_{0} \ll \dot{M}_{1}$, the thick accretion flow carries almost all the accretion rate and the ratio of the accretion rate through the thick accretion flow to that through the thin accretion disk is very large.
As $\dot{M}_{0}$ increases, that ratio gradually decreases and becomes about one around $\dot{M}_{0} \sim \dot{M}_{1}$.
It further decreases associated with the $\dot{M}_{0}$ increase and tends to converge to about a tenth when $\dot{M}_{0}$ exceeds $\dot{M}_{2}$.

\subsection{Effect of X-ray irradiation on the thick accretion flow}
We check an effect of the X-ray irradiation on a thick accretion flow, here.

The accretion rate through the thick accretion flow, $\dot{M}_{\rm ac,\; e}$, approximately relates with its half thickness, $h_{\rm ac,\; e}$, number density, $n_{\rm ac,\; e}$, and in-flow velocity, $u_{\rm e}$, as
\begin{equation}
\dot{M}_{\rm ac,\; e} \simeq 4\pi r h_{\rm ac,\; e} n_{\rm ac,\; e} m_{\rm p} u_{\rm e},
\label{eqn:Mdot_ac,e-nrh}
\end{equation}
while the irradiating X-ray luminosity is expressed as $L \simeq (\dot{M}_{0}/2) \eta c^{2}$.
From these equations, the ionization parameter of the thick accretion flow, $\xi_{\rm ac,\; e}$, is estimated as
\begin{eqnarray}
\xi_{\rm ac,\; e} &\simeq& 4\pi \eta m_{\rm p} c^{2} \alpha \left( \frac{\dot{M}_{0}/2}{\dot{M}_{\rm ac,\; e}}\right) \left( \frac{GM}{r}\right)^{-1} \left( \frac{kT_{\rm ac,\; e}}{m_{\rm p}}\right)^{3/2} \nonumber \\
&\simeq& 1.1 \times 10^{3} \left( \frac{\eta}{0.1}\right) \left( \frac{\alpha}{0.1}\right) \left( \frac{M}{10}\right)^{-1} \left(\frac{2\dot{M}_{\rm ac,\; e}/\dot{M}_{0}}{0.1}\right)^{-1} \left( \frac{r}{10^{11}}\right) \left( \frac{T}{10^{6}}\right)^{3/2} \mbox{ erg cm s}^{-1},
\label{eqn:xi_ac,e}
\end{eqnarray}
by employing such approximations in the standard accretion disk theory as
\begin{equation}
u_{\rm e} \simeq \alpha \left(\frac{h_{\rm ac,\; e}}{r}\right)^{2} \left(\frac{GM}{r}\right)^{1/2},
\label{eqn:u}
\end{equation}
and 
\begin{equation}
\frac{h_{\rm ac,\; e}}{r} \simeq \left( \frac{kT_{\rm ac,\; e}}{m_{\rm p}}\right)^{1/2} \left( \frac{GM}{r}\right)^{-1/2},
\label{eqn:h/r_ac.e}
\end{equation}
where $T_{\rm ac,\; e}$ is the temperature of the thick accretion flow.

The matter flowing from the envelope into the thick accretion flow should initially have $T_{\rm ac,\; e} \sim 10^{6}$ K when $\dot{M}_{0} \gtrsim \dot{M}_{2}$ as discussed in subsubsection \ref{HighAccretionRateCase}.
Since $2\dot{M}_{\rm ac,\; e}/\dot{M}_{0} \sim 0,1$ in that case, however, $\xi_{\rm ac,\; e}$ is as large as 10$^{3}$ erg cm s$^{-1}$ as seen from equation (\ref{eqn:xi_ac,e}) and the temperature is expected to go up to $\sim10^{7}$ K due to the X-ray heating.
When $\dot{M}_{0} \ll \dot{M}_{2}$, on the other hand, the temperature of the matter from the envelope should be $\sim 10^{7}$ K and the cooling time is considered to be longer than the matter inflowing time in this case.

Thus, we can expect that the thick accretion flow is kept thick with the temperature $\sim10^{7}$ K irrespectively of $\dot{M}_{0}$.

\subsection{Optical depth of the thick accretion flow}
We can roughly calculate an optical depth for the Thompson scattering in the radial direction of the thick accretion flow, $\tau$, as $\tau \simeq \sigma_{\rm T} n_{\rm ac,\; e} r$, where $\sigma_{\rm T}$ is the Thompson scattering cross section.
$n_{\rm ac,\; e}$ can be obtained from equation (\ref{eqn:Mdot_ac,e-nrh}) and $\dot{M}_{\rm ac,\; e}$ can be expressed as
\begin{equation}
\dot{M}_{\rm ac,\; e} =\left(\frac{\dot{M}_{\rm ac,\; e}}{\dot{M}_{0}/2}\right) \left(\frac{\dot{M}_{0}}{\dot{M}_{\rm E}}\right) \frac{2\pi m_{\rm p}GM}{ c \sigma_{\rm T}},
\label{eqn:M_ac,e-LE}
\end{equation}
where $\dot{M}_{\rm E}$ is the Eddington accretion rate defined as $\dot{M}_{\rm E} = 4\pi m_{\rm p} GM/(c \sigma_{\rm T}$).
Then, the result is
\begin{eqnarray}
\tau &\simeq& \frac{1}{2 \alpha c} \left( \frac{\dot{M}_{\rm ac,\; e}}{\dot{M}_{0}/2}\right) \left( \frac{\dot{M}_{0}}{\dot{M}_{\rm E}}\right) \left( \frac{GM}{r}\right)^{2} \left( \frac{kT_{\rm ac,\; e}}{m_{\rm p}}\right)^{-3/2} \nonumber \\ &\simeq& 0.1  \left( \frac{\alpha}{0.1}\right)^{-1} \left( \frac{2\dot{M}_{\rm ac,\; e}/\dot{M}_{0}}{0.1}\right) \left( \frac{\dot{M}_{0}}{\dot{M}_{\rm E}}\right) \left( \frac{M}{10}\right)^{2} \left( \frac{r}{10^{11}}\right)^{-2} \left( \frac{T_{\rm ac,\; e}}{10^{7}}\right)^{-3/2}.
\label{eqn:tau}
\end{eqnarray}
Thus, $\tau < 1$ is expected for $r \sim 10^{11}$ cm and $T_{\rm ac,\; e} \sim 10^{7}$ K unless $\dot{M}_{0}$ largely exceeds $\dot{M}_{\rm E}$, since $\dot{M}_{\rm ac,\; e}/(\dot{M}_{0}/2)$ becomes as small as 0.1 even when $\dot{M}_{0}$ is as large as $\dot{M}_{\rm E}$.

The temperature of the thick accretion flow is considered to get close to the virial temperature $\sim m_{\rm p}GM/(6kr)$ as $r$ decreases from $\sim 10^{11}$ cm.
In that case, the radial optical depth is approximately calculated as
\begin{eqnarray}
\tau &\simeq& \frac{2^{1/2}3^{3/2}}{ \alpha c} \left( \frac{\dot{M}_{\rm ac,\; e}}{\dot{M}_{0}/2}\right) \left( \frac{\dot{M}_{0}}{\dot{M}_{\rm E}}\right) \left( \frac{GM}{r}\right)^{1/2} \nonumber \\ 
&\simeq& 0.5  \left( \frac{\alpha}{0.1}\right)^{-1} \left( \frac{2\dot{M}_{\rm ac,\; e}/\dot{M}_{0}}{0.1}\right) \left( \frac{\dot{M}_{0}}{\dot{M}_{\rm E}}\right) \left(\frac{r}{100\ R_{\rm S}}\right)^{-1/2},
\label{eqn:tau_R_S}
\end{eqnarray}
where $R_{\rm S}$ is the Schwarzschild radius.
Thus, $\tau < 1$ could be ensured even in the inner region than $10^{11}$ cm except the X-ray emitting region with several 10 $R_{\rm S}$, as far as $\dot{M}_{0} \lesssim \dot{M}_{\rm E}$.

\section{Properties of excretion flows}
After sufficient angular momentum is transferred via the viscous stress in the core and the envelope, as discussed above, a thin accretion disk and a thin excretion disk extend from the core, and a thick accretion flow and a thick excretion flow spread out from the envelope, respectively.
We study properties of the thin excretion disk and the thick excretion flow in this section.

\subsection{Thin excretion disk}
We designate the mass inflow rate through the thin accretion disk and the mass outflow rate through the thin excretion disk as $\dot{M}_{\rm ac,\; c}$ and $\dot{M}_{\rm ex,\; c}$ respectively and assume $\dot{M}_{\rm ac,\; c} = \dot{M}_{\rm ex,\; c}$.
Then, the angular momentum of $\sim \dot{M}_{\rm ac,\; c}\sqrt{r_{0}GM}$ is considered to be transfered from the thin accretion disk to the thin excretion disk in the core, following the discussions in subsection \ref{OuterBoundary}.
Since the outflowing matter through the thin excretion disk has carried the angular momentum of $\sim \dot{M}_{\rm ex,\; c}\sqrt{r_{0}GM}$ from the stream, we can write an equation for the angular momentum transfer in the thin excretion disk as
\begin{eqnarray}
\dot{M}_{\rm ex,\; c} \sqrt{rGM} + 3\pi \nu \Sigma \sqrt{rGM} &\simeq& \dot{M}_{\rm ex,\; c} \sqrt{r_{0}GM} + \dot{M}_{\rm ac,\; c} \sqrt{r_{0}GM} \nonumber \\ &=& 2 \dot{M}_{\rm ex,\; c} \sqrt{r_{0}GM}.
\label{eqn:thin-AMoutflow_rate}
\end{eqnarray}
on an assumption that the matter at any places in the thin excretion disk is rotating with the respective Keplerian circular velocity.

The structures of the thin excretion disk are likely to be the same as those of the standard accretion disk except that the flow direction is opposite and that the angular momentum flow rate is much larger than that in the accretion disk.
The centrifugal force basically balances with the gravitational force in the radial direction, while the hydrostatic equilibrium is established in the direction perpendicular to the equatorial plane.
The energy generated by the viscous stress is radiated away from the surface of the disk.

As seen from equation (\ref{eqn:thin-AMoutflow_rate}), $\nu \Sigma$ should have a relation as
\begin{equation}
\nu \Sigma \simeq (\nu \Sigma)_{r=r{0}}\left( 2\sqrt{\frac{r_{0}}{r}}\; - \; 1\right).
\label{eqn:nu-variation}
\end{equation}
This indicates that the thin excretion disk cannot extend beyond an outermost radius, $r_{\rm out}$, which is given as
\begin{equation}
r_{\rm out} \simeq 4 r_{0}.
\label{eqn:r_out}
\end{equation}
Since the outermost radius exceeds the Roche lobe radius, however, the practical outer boundary of the thin excretion disk is likely to be the tidal radius slightly smaller than the Roche lobe radius (e.g. subsection 5.11 in Frank et al. 2002).

$\nu \Sigma$ decreases and thus the energy generation rate through the viscous stress lessens as $r$ increases as seen in equation (\ref{eqn:nu-variation}).
As a result, the disk temperature decreases and the disk thickness tends to be zero as $r$ increases.
Hence, the excretion disk is expected to slash through the stream without large interaction with it.

\subsection{Thick excretion flow}
Similarly to the case of the thin excretion disk, we designate the mass inflow rate through the thick accretion flow and the mass outflow rate through the thick excretion flow as $\dot{M}_{\rm ac,\; e}$ and $\dot{M}_{\rm ex,\; e}$ respectively and assume $\dot{M}_{\rm ac,\; e} = \dot{M}_{\rm ex,\; e}$.
Then, the angular momentum of $\sim \dot{M}_{\rm ac,\; e}\sqrt{r_{0}GM}$ is transfered from the thick accretion flow to the thick excretion flow in the envelope, and we have an equation for the angular momentum transfer in the thick excretion flow as
\begin{equation}
\dot{M}_{\rm ex,\; e} j_{\rm ex,\; e} - 2\pi r^{3} \nu \Sigma \frac{\partial \Omega}{\partial r} \simeq 2 \dot{M}_{\rm ex,\; e} \sqrt{r_{0}GM},
\label{eqn:thickAMoutflow_rate}
\end{equation}
where $j_{\rm ex,\; e}$ is the specific angular momentum carried by the outflowing matter.

Energy is also transfered via the viscous stress from the accretion flow to the excretion flow and the energy transfer rate is approximately given as 
\begin{equation}
-\left( 2\pi r^{3} \nu \Sigma \Omega \frac{\partial \Omega}{\partial r}\right)_{r = r_{0}} \simeq \dot{M}_{\rm ac,\; e} \frac{GM}{r_{0}} = \dot{M}_{\rm ex,\; e} \frac{GM}{r_{0}}.
\label{eqn:EFR_r_0}
\end{equation}
Thus, the specific energy of the outflowing matter can be expressed as
\begin{equation}
\frac{v^{2}}{2} + \frac{5kT}{m_{\rm p}} - \frac{GM}{r}  - \frac{2\pi r^{3} \nu \Sigma}{\dot{M}_{\rm ex,\; e}} \Omega \frac{\partial \Omega}{\partial r} \simeq -\frac{1}{2} \frac{GM}{r_{\rm L1}} + \frac{GM}{r_{0}},
\label{eqn:Bernoulli_out}
\end{equation}
by adding the energy transferred from the inflowing accretion flow to that carried from the stream as given in equation (\ref{eqn:Bernoulli}).
Here, it is assumed that the radiative cooling is negligible in this thick flow.
Since the energy of the outflowing matter is positive as seen from the right side of the above equation, the thick excretion flow does not stop at $r_{\rm out}$ in this case but rather tends to be a wind flow having the constant specific angular momentum $j_{\rm ex,\; \infty} = 2 \sqrt{r_{0}GM}$ and the specific energy given as 
\begin{equation}
\frac{v^{2}}{2} + \frac{5kT}{m_{\rm p}} - \frac{GM}{r}  \simeq -\frac{1}{2} \frac{GM}{r_{\rm L1}} + \frac{GM}{r_{0}}.
\label{eqn:Bernoulli_out}
\end{equation}
This flow can go out to the infinity with the terminal velocity slightly less than $\sqrt{2GM/r_{0}} \sim 10^{3}$ km s$^{-1}$ for $M \sim 10 M_{\odot}$ and $r_{0} \sim 10^{11}$ cm.


\section{Summary and discussion}
Properties of an accretion ring and those of accretion and excretion flows from it in the steady state have been studied.

An initial situation is thought that matter inflowing from a companion star in a close binary with a compact object forms a circularly rotating ring along the Keplerian circular orbit determined by the average specific angular momentum carried with the inflowing matter around the compact object.
The accretion ring is defined as the sojourning place of the matter inflowing from the companion star where the specific angular momentum distribution changes from the initial constant one to the final one as $\sqrt{rGM}$ to spread out to accretion and excretion flows.

The studies in the relevant literatures have, however, estimated the life time of the accretion ring very short neglecting the time to change the specific angular momentum distribution and paid no particular attention to the accretion ring.

This paper shows that the time for the change of the angular momentum distribution is not negligibly short but rather long so as for the accretion ring to play several important roles in accretion and excretion processes.

\subsection{Two-layer accretion flow}
If the total energy of the inflowing matter to the ring is conserved, the matter should form a thick envelope around the ring.
Then, two internal flows are considered to appear in the thick envelope.  
One is an internal flow separating to thick accretion and excretion flows as a result of the angular momentum transfer in the envelope.
The other is a cooling flow towards the envelope center via the radiative cooling under an effect of X-ray irradiation.
This flow is expected to form a core around the torus center from which  thin accretion and excretion disks extend.

Evaluating the time scales for the two internal flows respectively and comparing them with each other, the properties of the internal flows are classified into the following three cases in terms of the intrinsic mass flow rate from the companion star, $\dot{M}_{0}$.
\begin{itemize}
\item Low accretion rate case
\end{itemize}
When $\dot{M}_{0}$ is much less than $\dot{M}_{1}$ in equation (\ref{eqn:M_1}), the density of the ring-tube is too low for the matter to cool down and the ring-tube is kept thick with the initial temperature, $T_{\rm in}$.
\begin{itemize}
\item Medium accretion rate case
\end{itemize}
When $\dot{M}_{0}$ is as large as or larger than $\dot{M}_{1}$ but much less than $\dot{M}_{2}$ defined in equation (\ref{eqn:Mdot_2}), the effect of the X-ray heating is not significant but two parallel accretion flows are expected, a thin accretion disk from the core and a thick accretion flow from the envelope.
The two parallel flows have the comparable accretion rate to each other but that in the thin disk tends to be dominant to the other as $\dot{M}_{0}$ gets close to $\dot{M}_{2}$.
\begin{itemize}
\item High accretion rate case
\end{itemize}
When $\dot{M}_{0}$ is as large as or larger than $\dot{M}_{2}$, the effect of the X-ray heating is significant.
A two-layer accretion flow is expected in this case too but the accretion rate through the thin disk is largely dominant to that through the thick flow.

The above study predicts that the accretion ring generally originates a two-layer accretion flow in which a thin disk is sandwiched by a thick flow, unless the accretion rate is very low.
This prediction is consistent with the following discussion from observations.

Churazov, Gilfanov \& Revnivtsev (2001), analyzing data of the typical black hole source, Cyg X-1, argues that the overall shape of the power density spectra can be qualitatively explained if a geometrically thin disk responsible for the relatively stable soft component is sandwiched by a geometrically thick flow, responsible for the variable hard component, extending up to a large distance from the compact object.
Sugimoto et al. (2016) obtained the power density spectra of Cyg X-1 on the lower frequency side than Churazov, Gilfanov \& Revnivtsev analyzed and show that the same features of the power density spectra extends to the frequencies as low as 10$^{-6} \sim 10^{-7}$ Hz.
Thus, if the two-layer flow is really responsible for the extension of the characteristic power density spectra down to such a very low frequency, it is required to start from almost the farthest end of the accretion flow (Inoue 2021a).

\subsection{Thick ring-envelope}
The present study also predicts that the accretion ring commonly has a thick envelope.

Several low-mass X-ray binaries (LMXBs) are known to repeatedly exhibit absorption dips 
in the X-ray light curves with or close to their respective orbital periods.  
Diaz Trigo et al. (2006) investigated spectral changes over the binary period as well as during dips in six bright 
dipping LMXBs in detail.  They found presence of an ionized plasma having a cylindrical 
geometry, since the absorption line properties do not vary strongly with orbital phase 
outside the dips.  This picture is consistent with presence of the hot 
ring-envelope in those dipping sources.  Since dipping sources are considered to be normal 
LMXBs viewed with large inclination angles, such ring-envelopes should 
be common to LMXBs.  

It is discussed in this study that the ionization parameter, $\xi$, and the temperature, $T$, of the ring-envelope should be around 10$^{2}$ erg cm s$^{-1}$ and 10$^{6}$ K respectively in the case of the high accretion rate.

Inoue (2019) obtained the $\xi$ and $T$ values from the best fit parameters in the reproductions the observed light curves of three bright X-ray pulsars with the precessing accretion ring model originally proposed by Inoue (2012). 
Those values roughly agree to the above discussions.

\subsection{Thick excretion flow}
Thick excretion flows are predicted to come out from the thick ring-envelopes as well.

This thick excretion flow could be a likely origin of absorption lines of highly ionized (He and H-like) Fe observed from several binary X-ray sources.
Those sources include black hole binaries (e.g. Ueda et al. 1998; Kotani et al. 2000), and weakly magnetized neutron star binaries (e.g. Diaz Trigo et al. 2006).
High resolution spectroscopy of absorption line features of highly ionized ions observed from the neutron star binary, GX13+1 revealed that the plasma responsible for the absorption lines is outflowing with a velocity of $\sim$400 km s$^{-1}$ (Ueda et al. 2004).

\subsection{Thin excretion disk}
This study further predicts that thin excretion disks extend from the ring-cores.

Although several discussions have been done on the outer edge of the disk close to the Roche lobe radius with particular interests on activities in the cataclysmic variables (e.g. subsection 5.11 in Frank et al. 2002), no clear observational evidence has been found to clarify natures of the outer edge of the disk in X-ray binaries.
It could be because the thermal emission from the excretion disk is considered to decrease as its radial position gets farther from the ring and the outer boundary could be disturbed by tides from the companion star.

We note an interesting observational evidence from active galactic nuclei (AGN) that the ratio of the innermost radius of the dust torus to the radius of the broad line region is about 4 (e.g. Koshida et al. 2014).
Inoue (2021a) discusses that some properties of the broad line region are similar to those of accretion ring.
If the broad line region really corresponds to the accretion ring, an excretion disk is expected to extend and to form another ring with a radius 4 times as large as that of the broad line region.
This discussion tempts us to consider that the innermost radius of the dust torus could relate to the outer boundary of the excretion disk.
There could also exist a possibility for the thick excretion flow to relate with the bi-polar cones often observed from AGNs (e.g. H\"{o}nig 2019).
These possibilities are discussed separately (Inoue 2021b).\\

As studied in this paper, accretion rings are likely to play some important roles in accretion flows onto compact objects in X-ray binaries.
Since a number of assumptions, approximations and simplifications have been employed in the present study, 
further and more detailed studies on the accretion rings are anticipated.


\begin{ack}
The main part of this study was once submitted to this journal but was rejected.  Comments from the referee on the original manuscript let the author notice that there had remained several important points to discuss, and were very helpful to re-submit the present manuscript.  The author is grateful to the previous referee for the critical comments.
\end{ack}

\appendix 
\section*{Effect of X-ray heating on the cooling time}
Let us express the radiative cooling rate per volume and the X-ray heating rate per volume by introducing the cooling function $\Lambda (T)$ and the heating function $\Phi (T)$ as $-n^{2} \ \Lambda (T)$ and 
$(L/r_{0}^{2}) n\ \Phi (T)$ respectively.
Then, the time evolution of the thermal energy density, $E = 3nkT$, is given as
\begin{equation}
\frac{dE}{dt} \  =\  \frac{L}{r_{0}^{2}} n \Phi (T) \ - \  n^{2} \Lambda (T).
\label{eqn:dEdt}
\end{equation}
If the heating and cooling rates balance with each other when $L = L_{\rm eq}$ and $n = n_{\rm eq}$, we can have a relation as
\begin{equation}
\frac{\Lambda (T)}{\Phi (T)} = \frac{L_{\rm eq}}{n_{\rm eq}r_{0}^{2}} = \xi_{\rm eq} (T).
\label{eqn:Lambda/Phi}
\end{equation}
With this equation, equation (\ref{eqn:dEdt}) can be rewritten as
\begin{equation}
\frac{dE}{dt} \ =\ - \ n^{2} \Lambda (T) \ \left( 1 \ - \ \frac {\xi (T)}{\xi_{\rm eq}(T)} \right).
\label{eqn:dEdt_R}
\end{equation}
From this equation, we can calculate the cooling time, $t_{\rm c}$, as
\begin{equation}
t_{\rm c} \simeq \frac{E}{|dE/dt|} = \varepsilon \ t_{\rm c,\; 0},
\label{eqn:t_c}
\end{equation}
where $t_{\rm c,\; 0}$ is the cooling time when no X-ray heating exists and given as 
\begin{equation}
t_{\rm c,\; 0} \simeq \frac{3kT}{n\Lambda (T)},
\label{eqn:t_c,0}
\end{equation}
and $\varepsilon$ is the parameter expressing the effect of X-ray heating, defined as
\begin{equation}
\varepsilon \ = \ \left( 1 \ - \ \frac {\xi (T)}{\xi_{\rm eq}(T)} \right)^{-1}.
\label{varepsilon}
\end{equation}


\end{document}